\documentclass[conference]{IEEEtran}

\usepackage{graphicx}
\usepackage{amsmath}
\usepackage{amssymb}
\usepackage{color}
\usepackage{ifpdf}
\usepackage{float}
\usepackage[utf8]{inputenc}
\usepackage{multirow}
\usepackage{rotating}
\usepackage[caption=false]{subfig}

\usepackage{moresize}
\usepackage[hyphens]{url}
\usepackage{booktabs}
\usepackage{listings}
\usepackage{paralist}
\usepackage{wrapfig}
\usepackage{multirow}
\usepackage{ifpdf}
\usepackage{xspace}
\usepackage{keyval}
\usepackage{color}

\usepackage{comment}

\definecolor{listinggray}{gray}{0.95}
\definecolor{darkgray}{gray}{0.7}
\definecolor{commentgreen}{rgb}{0, 0.4, 0}
\definecolor{darkblue}{rgb}{0, 0, 0.4}
\definecolor{middleblue}{rgb}{0, 0, 0.7}
\definecolor{darkred}{rgb}{0.4, 0, 0}
\definecolor{brown}{rgb}{0.5, 0.5, 0}
\definecolor{dkgreen}{rgb}{0,0.5,0}
\definecolor{orange}{rgb}{1,.5,0}
\definecolor{dandelion}{cmyk}{0,0.29,0.84,0}

\usepackage[normalem]{ulem}
\makeatletter
\def\cyanuwave{\bgroup \markoverwith{\lower3.5\p@\hbox{\sixly \textcolor{cyan}{\char58}}}\ULon}
\def\reduwave{\bgroup \markoverwith{\lower3.5\p@\hbox{\sixly \textcolor{red}{\char58}}}\ULon}
\def\blueuwave{\bgroup \markoverwith{\lower3.5\p@\hbox{\sixly \textcolor{blue}{\char58}}}\ULon}
\font\sixly=lasy6 
\makeatother

\let\labelindent\relax
\usepackage{enumitem}

\newif\ifdraft
\ifdraft
 \newcommand{\N}[1]{\textbf{*** NOTE: #1}\xspace}
 \newcommand{\jhanote}[1]{ {\textcolor{red} { ***SJ: #1 }}}
 \newcommand{\mtnote}[1]{ {\textcolor{orange} { ***MT: #1 }}}
 \newcommand{\note}[1]{ {\textcolor{brown} { *** #1 }}}
 \newcommand{\jdnote}[1]{ {\textcolor{cyan} { ***JD: #1 }}}
 \newcommand{\dwwnote}[1]{ {\textcolor{blue} { ***DWW: #1 }}}
\else
 \newcommand{\N}[1]{}
 \newcommand{\jhanote}[1]{}
 \newcommand{\mtnote}[1]{}
 \newcommand{\jdnote}[1]{}
 \newcommand{\dwwnote}[1]{}
 \newcommand{\note}[1]{}
\fi

\lstdefinestyle{myListing}{
  frame=single,
  backgroundcolor=\color{listinggray},
  language=C,
  basicstyle=\ttfamily \footnotesize,
  breakautoindent=true,
  breaklines=true
  tabsize=2,
  captionpos=b,
  aboveskip=0em,
  belowskip=-2em,
}

\lstdefinestyle{myPythonListing}{
  frame=single,
  backgroundcolor=\color{listinggray},
  language=Python,
  basicstyle=\ttfamily \scriptsize,
  breakautoindent=true,
  breaklines=true
  tabsize=2,
  captionpos=b,
}

\ifpdf
\DeclareGraphicsExtensions{.pdf, .jpg, .tif}
\else
\DeclareGraphicsExtensions{.ps,  .eps, .jpg}
\fi

\begin{document}

\title{High-throughput Binding Affinity Calculations at Extreme Scales}

\author{
    \IEEEauthorblockA{
      Jumana Dakka$^=$\IEEEauthorrefmark{1},
        Matteo Turilli$^=$\IEEEauthorrefmark{1},
        David W Wright$^=$\IEEEauthorrefmark{2},
        Stefan J Zasada$^=$\IEEEauthorrefmark{2},\\
        Vivek Balasubramanian\IEEEauthorrefmark{1},
        Shunzhou Wan\IEEEauthorrefmark{2},
        Peter V Coveney\IEEEauthorrefmark{2} and
        Shantenu Jha\IEEEauthorrefmark{1},\IEEEauthorrefmark{3},\IEEEauthorrefmark{4}
    }
    \IEEEauthorblockA{
       \IEEEauthorrefmark{1}Department of Electric and Computer Engineering, Rutgers University, NJ, USA
    }
    \IEEEauthorblockA{
        \IEEEauthorrefmark{2}Centre for Computational Sciences, UCL
    }
    \IEEEauthorblockA{
        \IEEEauthorrefmark{3} Institute for Advanced Computational Sciences, Stony Brook University, NY, USA
    }

    \IEEEauthorblockA{
        \IEEEauthorrefmark{4}Computational Science Initiative, Brookhaven National Laboratory
        }
    $^=$ First authors, contributed equally, alphabetical order
 }

\maketitle

\begin{abstract}
Resistance to chemotherapy and molecularly targeted therapies is a major
factor in limiting the effectiveness of cancer treatment. In many cases,
resistance can be linked to genetic changes in target proteins, either
pre-existing or evolutionarily selected during treatment. Key to overcoming
this challenge is an understanding of the molecular determinants of drug
binding. Using multi-stage pipelines of molecular simulations we can gain
insights into the binding free energy and the residence time of a ligand,
which can inform both stratified and personal treatment regimes and drug
development.
To support the scalable, adaptive and automated calculation of the binding
free energy on high-performance computing resources, we introduce the
High-throughput Binding Affinity Calculator (HTBAC). HTBAC uses a building
block approach in order to attain both workflow flexibility and performance.
We demonstrate close to perfect weak scaling to hundreds of concurrent
multi-stage binding affinity calculation pipelines. This permits a rapid
time-to-solution that is essentially invariant of the calculation protocol,
size of candidate ligands and number of ensemble simulations. 
As such, HTBAC advances the state of the art of binding affinity calculations 
and protocols.
HTBAC provides the platform to enable scientists to study a wide range of cancer drugs and
candidate ligands in order to support personalized clinical decision making
based on genome sequencing and drug discovery.

\end{abstract}

\section{Introduction}\label{sec:intro}
In recent years, chemotherapy based on targeted kinase inhibitors (TKIs) has
played an increasingly prominent role in the treatment of cancer. TKIs have
been developed to selectively inhibit kinases involved in the signaling
pathways that control growth and proliferation, which often become
dysregulated in cancers. This targeting of specific cancers reduces the risk
of damage to healthy cells and increases treatment success. Currently, 35
FDA-approved small molecule TKIs are in clinical use, and they represent a
significant fraction of the \$37 billion U.S. market for oncology
drugs~\cite{FDA, Zhao2014}. Imatinib, the first of these of drugs, is
partially credited for doubling survivorship rates in certain
cancers~\cite{Zhao2014, ACSreport}.

Unfortunately, the development of resistance to these drugs limits the amount
of time that patients can derive benefits from their treatment. Resistance to
therapeutics is responsible for more than 90\% of deaths in patients with
metastatic cancer~\cite{Longley2005}. While drug resistance can emerge via
multiple mechanisms, small changes to the chemical composition of the
therapeutic target (known as mutations) control treatment sensitivity and
drive drug resistance in many patients (see Fig.~\ref{fig:egfr}). In some
commonly targeted kinases such as Abl, these changes account for as many as
90\% of treatment failure~\cite{Shah2002}.

There are two major strategies for countering the threat to treatment
efficacy posed by resistance: tailoring the drug regimen received by a
patient according to the mutations present in their particular cancer, and
developing more advanced therapies that retain potency for known resistance
mutations. In both cases, future developments require insight into the
molecular changes produced by mutations, as well as ways to predict their
impact on drug binding on a timescale much shorter than is typically
experimentally feasible. This represents a grand challenge for computational
approaches.

The rapidly decreasing cost of next-generation sequencing has led many cancer
centers to begin deep sequencing of patient tumors to identify the genetic
alterations driving individual cancers. The ultimate goal is to make
individualized therapeutic decisions based upon these data---an approach
termed \textit{precision cancer therapy}. While several common (recurrent)
mutations have been cataloged for their ability to induce resistance or for
their susceptibility to particular kinase inhibitors, the vast majority of
clinically observed mutations are rare. Essentially, this ensures that it
will be impossible to make therapeutic decisions about the majority of
individual patient tumors by using catalog-building alone.

Fortunately, concurrent improvements in computational power and algorithm
design are enabling the use of molecular simulations to reliably quantify
differences in binding strength. This provides the opportunity to use
advances in molecular simulations to supplement existing inductive decision
support systems with deductive predictive modeling and drug
ranking~\cite{Marias2011, Sloot2009}. Where existing systems based on
statistical inference are inherently limited in their range of applicability
by the existence of data from previous similar cases, the addition of
modeling allows evidence based decision making even in the absence of direct
past experience.

The same molecular simulation technologies that can be employed to
investigate the origins of drug resistance can also be used to design new
therapeutics. Creating simulation protocols which have well defined
uncertainty and produce statistically meaningful results represents a
significant computational challenge. Furthermore, it is highly likely that
differences among investigated systems will demand different protocols as
studies progress. For example, drug design programmes often require the rapid
screening of thousands of candidate compounds to filter out the worst binders
before using more sensitive methods to refine the structure. Not all changes
induced in protein shape or behavior are local to the drug binding site and,
in some cases, simulation protocols will need to adjust to account for
complex interactions between drugs and their targets within individual
studies.

Recent work that used molecular simulations to provide input to machine
learning models~\cite{Ash2017} required simulations of 87 compounds even if
they were designed merely to distinguish the highly active from weak
inhibitors of the ERK2 kinase. If we wish to build on such studies to help
inform later stages of the drug discovery pipeline, in which much more subtle
alterations are involved, it is likely a much larger number of simulations
will be required. This is before we begin to consider the influence of
mutations or the selectivity of drugs to the more than 500 different genes in
the human kinome~\cite{Li2016}.

For molecular simulations to make the necessary impact, the dual challenge of
scale (thousands of concurrent multi-stage pipelines) and sophistication
(adaptive selection of binding affinity protocols based upon statistical
errors and uncertainty) will need to be tackled. Tools that facilitate the
scalable and automated computation of varied binding free energy calculations
on high-performance computing resources are necessary. To achieve that goal,
we introduce the High-throughput Binding Affinity Calculator~(HTBAC). HTBAC
applies recent advances in workflow system building blocks to the accurate
calculation of binding affinities, executing hundreds of concurrent
calculations on a leadership class machine. This permits the rapid time-to-
solution that is essentially invariant of the size of candidate ligands as
well as the type and number of protocols concurrently employed.


In the next Section, we provide details of ensemble molecular dynamics
approach and its advantages over the single trajectory approach. We also
introduce the ESMACS and related protocols to compute binding affinities
using ensemble-based approaches. In Section~\ref{sec:3}, we discuss the
computational challenges associated with the scalable execution of multiple,
and possibly concurrently executing protocols. Section~\ref{sec:4} introduces
RADICAL-Cybertools---a suite of building blocks to address the challenges
outlined in Section~\ref{sec:3}---and describe how they are used by HTBAC to
manage the execution of binding affinity calculations at extreme scales.
Experiments to characterize the performance overheads of RADICAL-Cybertools
and the weak scaling properties of the HTBAC implementation of the ESMACS
protocol on the Blue Waters supercomputer are discussed in
Section~\ref{sec:6}. 
We conclude with a discussion of the impact of HTBAC, implication for binding
affinity calculations and near-term future work.

\section{Methodology}\label{sec:meth}
The strength of drug binding is determined by a thermodynamic property known
as the binding free energy (or binding affinity). One promising technology for
estimating binding free energies and the influence of protein composition on
them is molecular dynamics (MD)~\cite{Karplus2005}. Our previous work
\cite{Sadiq2010, Wan2011} has demonstrated that running multiple MD
simulations based on the same system and varying only in initial velocities
offers a highly efficient method of obtaining accurate and reproducible
estimates of the binding affinity. We term this approach ensemble molecular
dynamics, ``ensemble'' here referring to the set of individual (replica)
simulations conducted for the same physical system. In this Section we discuss
the advantages to this approach.

\subsection{Ensemble Molecular Dynamics Simulations}

Atomistically detailed models of the drug and target protein can be used as the
starting point for MD simulations to
study the influence of mutations on drug binding. The chemistry of the system
is encoded in what is known as a potential~\cite{Karplus2002}. In the
parameterization of the models, each atom is assigned a mass and a charge,
with the chemical bonds between them modeled as springs with varying
stiffness. Using Newtonian mechanics the dynamics of the protein and drug can
be followed and, using the principles of statistical mechanics, estimates of
thermodynamic properties obtained from simulations of single particles.

The potentials used in the simulations are completely under the control of
the user. This allows the user to manipulate the system in ways which would
not be possible in experiments. A particularly powerful example of this are
the so called ``alchemical'' simulations in which the potential used in the
simulation changes, from representing a particular starting system into one 
describing a related target system during execution. This allows for the 
calculation of free energy differences between the two systems, such as those 
induced by a protein mutation.

MD simulations can reveal how interactions change as a result of mutations,
and account for the molecular basis of drug efficacy. This understanding can
form the basis for structure-based drug design as well as helping to target
existing therapies based on protein composition. However, correctly capturing
the system behavior poses at least two major challenges: The approximations
made in the potential must be accurate enough representations of the real
system chemistry; and sufficient sampling of phase space is also required.

In order for MD simulations to be used as part of clinical decision support
systems, it is necessary that results can be obtained in a timely fashion.
Typically, interventions are made on a timescale of a few days or, at most, a
week. The necessity for rapid turn around times places additional demands on
simulation protocols which need to be optimized to gain results with a short
turn around time. Further to the scientific and practical considerations
outlined above, it is vital that reliable uncertainty estimates are
provided alongside all quantitative results for simulations to provide
actionable predictions.

We have developed a number of free energy calculation protocols based on the
use of ensemble molecular dynamics simulations with the aim of meeting these
requirements~\cite{Sadiq2008, Sadiq2010, Wan2017brd4, Wan2017trk}. Basing
these computations on the direct calculation of ensemble averages facilitates
the determination of statistically meaningful results along with complete
control of errors. The use of the ensemble approaches however, necessitates
the use of middleware to provide reliable coordination and distribution
mechanisms with low performance overheads.


\subsection{Protocols for Binding Affinity Calculations}

We have demonstrated the lack of reproducibility of single trajectory
approaches in both HIV-1 protease and MHC systems, with calculations for the
same protein-ligand combination, with identical initial structure and force
field, shown to produce binding affinities varying by up to 12 kcal mol
$^{-1}$ for small ligands (flexible ligands can vary even
more)~\cite{Wan2015, Sadiq2010, Wright2014}. Indeed, our work has revealed
how completely unreliable single simulation based approaches are.

Our work using ensemble simulations have also reliably produced results in
agreement with previously published experimental findings~\cite{Sadiq2010,
Wan2011, Wright2014, Bhati2017, Wan2017brd4, Wan2017trk}, and correctly
predicted the results of experimental studies performed by colleagues in
collaboration~\cite{Bunney2015}. While the accuracy of force fields could be a
source of error, we know from our work to date that the very large
fluctuations in trajectory-based calculations account for the lion’s share of
the variance (hence also uncertainty) of the results.



We designed two free energy calculation protocols with the demands of clinical
decision support and drug design applications in mind: ESMACS (enhanced
sampling of molecular dynamics with approximation of continuum
solvent)~\cite{Wan2017brd4} and TIES (thermodynamic integration with enhanced
sampling)~\cite{Bhati2017}. The former protocol is based on variants of the
molecular mechanics Poisson-Boltzmann surface area (MMPBSA) end-point method,
the latter on the `alchemical' thermodynamic integration (TI) approach. In
both cases, ensembles of MD simulations are employed to perform averaging and
to obtain tight control of error bounds in our estimates. In addition, the
ability to run replica simulations concurrently means that, as long as
sufficient compute resources are available, turn around times can be
significantly reduced compared to the generation of single long trajectories.
The common philosophy behind the two protocols entails similar middleware
requirements: In this work we focus on the ESMACS protocol but all results are
applicable also to TIES.

Each replica within the ESMACS protocol consists of a series of simulation
steps followed by post production analysis. Generally, an ESMACS replica will
contain between 3 and 12 equilibration simulation steps followed by a
production MD run, all of which are conducted in the NAMD
package~\cite{Phillips2005}. The first step is system minimization, the
following steps involve the gradual release of positional constraints upon
the structure and the heating to a physiologically realistic temperature.
Upon completion of the MD simulation, free energy computation (via MMPBSA and
potentially normal mode analysis) is performed using
AmberTools~\cite{Case2005, MillerIII2012}.

The ESMACS protocol is highly customizable. Both the number of simulation replicas in the ensemble and the lengths of their runs can be varied to
obtain optimal performance for any given system. Using replicas that only
vary in the initial velocities assigned to the atoms of the system we have
defined a standard protocol which prescribes a 25 replica ensemble, each run
consisting of 2 ns of equilibration and 4 ns of production simulation. Our
protocol has produced bootstrap errors of below 1.5 kcal mol$^{-1}$ (despite
replica values varying by more than 10 kcal mol$^{-1}$) for a varied range of
systems including small molecules bound to kinases and more flexible peptide
ligands binding to MHC proteins \cite{Wan2015, Wright2014, Wan2017brd4}.
In these systems, the error we obtained more than halves between ensembles of
5 and 25 replicas but increases in ensemble size have generally produced only
small improvements. More generally though, there may be cases where it is
important to increase the sampling of phase space either through expanding
the ensemble or by considering multiple initial configurations.

The ESMACS protocol can also be extended to account for adaptation energies
involved in altering the conformation of the protein or ligand during
binding. Almost all MMPBSA studies in the literature use the so-called
1-trajectory method, in which the energies of protein-inhibitor complexes,
receptor proteins and ligands are extracted from the MD trajectories of the
complexes alone. The ESMACS protocol can additionally use separate ligand and
receptor trajectories to account for adaptation energies, providing further
motivation to deploy the protocol via flexible and scalable middleware.

\subsection{Benchmark kinase system}

\begin{figure}
  \centering
  \includegraphics[width=0.60\columnwidth]{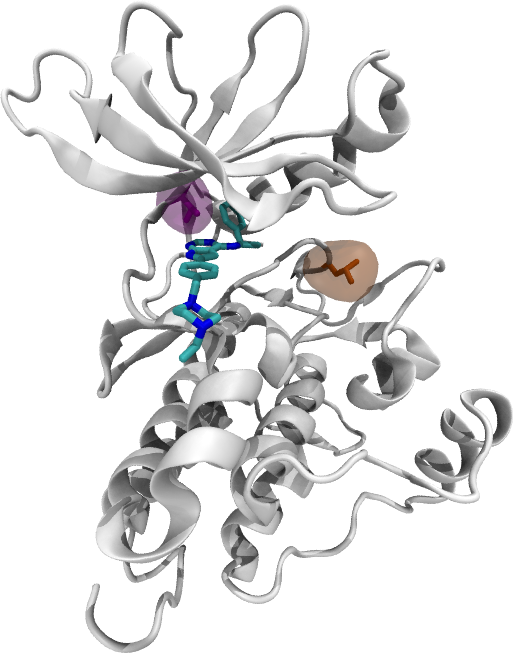}
  \caption{Cartoon representation of the EGFR kinase bound to the inhibitor
  AEE788 shown in chemical representation (based on PDB:2J6M). Two residues
  implicated in modulating drug efficacy are highlights; in pink T790 and in
  orange L858. Mutations to either of these residues significantly alter the
  sensitivity to TKIs.}\label{fig:egfr}
\end{figure}

A common target of kinase inhibitors is the epidermal growth factor receptor
(EGFR) which regulates important cellular processes including proliferation,
differentiation and apoptosis. EGFR is frequently over expressed in a range of
cancers, and is associated with disease progression and treatment. Clinical
studies have shown that EGFR mutations confer tumor sensitivity to tyrosine
kinase inhibitors in patients with non-small-cell lung cancer (examples shown
in Fig.~\ref{fig:egfr}) The tyrosine kinase domain of EGFR contains 288
residues, the full simulation system including solvent and the AEE788
inhibitor contains approximately 50 thousand atoms. The well established AMBER
ff99SBildn and GAFF force fields~\cite{Maier2015, Wang2004} were used to
parameterize the system for this work.

\subsection{Automated binding affinity calculations}

The implementation of any physically realistic molecular simulation has
always been an involved and multistage process, often requiring the scientist
to overcome a large manual overhead in the construction, preparation, and
execution protocols necessary to complete a set of simulations as well as to
invoke various analysis protocols for determining desired properties
post-production.

Several tools have been been developed to automate MD workflows for the
rapid, accurate and reproducible computation of binding free energies of
small molecules to their target proteins. For example, BAC~\cite{Sadiq2008}
is a partially automated workflow system which comprises (a) model building
(including incorporation of mutations into patient specific protein models),
(b) running ensembles of MD simulations using a range of free energy
techniques and (c) statistical analysis. In Section~\ref{sec:4}, we decribed
how we have enhanced BAC using the RADICAL-Cybertools to produce (HTBAC).

\section{Computational Challenges at Extreme Scale}\label{sec:3}
High-performance computing (HPC) environments were designed to primarily
support the execution of single simulations. Current HPC platforms enable the
strong and weak scaling of single tasks (hitherto mostly simulations), with
limited software and systems support for the concurrent execution of multiple
heterogeneous tasks as part of a single application (or workflow). As the
nature of scientific inquiry and the applications to support that inquiry
evolve, there is a critical need to support the scalable and concurrent
execution of a large number of heterogeneous tasks.

Sets of tasks with dependencies that determine the order of their execution
are usually referred to as ``workflows''. Often times, the structure of the
task dependencies is simple and adheres to discernible patterns, even though
the individual tasks and their duration are non-trivially distinct. Put
together, it is a challenge to support the scalable execution of workflows on
HPC resources due to the existing software ecosystem and runtime systems
typically found.

Many workflow systems have emerged in response to the aforementioned problem.
Each workflow system has its strengths and unique capability, however each
system typically introduces its problems and challenges. In spite of the many
successes of workflow systems, there is a perceived high barrier-to-entry,
integration overhead and limited flexibility.

Interestingly, many commonly used workflow systems in high-performance and
distributed computing emerged from an era when the software landscape
supporting distributed computing was fragile, missing features and services.
Not surprisingly, initial workflow systems had a monolithic design that
included the end-to-end capabilities needed to execute workflows on
heterogeneous and distributed cyberinfrastructures. Further, these workflow
systems were typically designed by a set of specialists to support large
``big science'' projects such as those carried out at the
LHC~\cite{breskin2009cern} or LIGO~\cite{althouse1992ligo}. The fact that the
same workflow would be used by thousands of scientists over many years
justified, if not amortized, the large overhead of integrating application
workflows with monolithic workflow systems. This influenced the design and
implementation of interfaces and programming models.

However as the nature, number and usage of workflows has evolved so have the
requirements: scale remains important but only when delivered with the
ability to prototype quickly and flexibly. Furthermore, there are also new
performance requirements that arise from the need to support concurrent
execution of heterogeneous tasks. For example, when executing multiple
homogeneous pipelines of heterogeneous tasks, for reasons of efficient
resource utilization there is a need to ensure that the individual pipelines
have similar execution times. The pipeline-to-pipeline fluctuation must be
minimal while also managing the task-to-task runtime fluctuation across
concurrently executing pipelines.


Thus the flexible execution of heterogeneous ensembles MD simulations face
both system software and middleware challenges: existing system software that
is typically designed to support the execution of single large simulations on
the one hand, and workflow systems that are designed to support specific use
cases or `locked-in' end-to-end executions. In the next Section, we discuss
the design and implementation of the RADICAL-Cybertools, a set of software
building blocks that can be easily composed to design, implement and execute
domain specific workflows rapidly and at scale.

\section{RADICAL-Cybertools}\label{sec:4}
We have designed RADICAL-Cybertools (RCT) to be functionally delineated
middleware building blocks and to address some of the challenges in
developing and executing workflows on HPC platforms. HTBAC uses two RCT
components, mainly the Ensemble Toolkit (EnTK) and RADICAL-Pilot (RP).
EnTK provides the ability to create and execute ensemble-based
workflows/applications with complex coordination and communication but
without the need for explicit resource management. EnTK uses RP as a
runtime system which provides resource management and task execution
capabilities.

RCT eschew the concept of a monolithic workflow systems and uses ``building
blocks''. RCT provide scalable implementations of building blocks in Python
that are used to support dozens of scientific applications on
high-performance and distributed systems~\cite{turilli2016analysis,angius2017converging,treikalis2016repex, balasubramanian2016ensemble,balasubramanian2016extasy}. In this Section we discuss details of RP, EnTK
and HTBAC, understanding how these components have been used to support the
flexible and scalable execution of pipelines.

\subsection{RADICAL-Pilot}

The scalable execution of applications with large ensembles of tasks is
challenging. Traditionally, two methods are used to execute multiple tasks on
a resource: each task is scheduled as an individual job, or message-passing
interface (MPI) capabilities
are used to execute multiple tasks as part of a single job. The former method
suffers from unpredictable queue time: each task independently awaits in the
resource's queue to be scheduled. The latter method relies on the fault
tolerance of MPI, and is suitable to execute tasks that are homogeneous and
have no interdependencies.

The Pilot abstraction~\cite{turilli2017comprehensive} solves these issues:
The pilot abstraction: (i) uses a placeholder job without any tasks assigned
to it, so as to acquire resources via the local resource management system
(LRMS); and, (ii) decouples the initial resource acquisition from
task-to-resource assignment. Once the pilot (container-job) is scheduled via
the LRMS, it can pull computational tasks for execution. This functionality
allows all the computational tasks to be executed directly on the resources,
without being queued via the LRMS\@. The pilot abstraction thus supports the
requirements of task-level parallelism and high-throughput as needed by
science drivers, without affecting or circumventing the queue policies of HPC
resources.

RADICAL-Pilot is an implementation of the pilot abstraction, engineered to
support scalable and efficient launching of heterogeneous tasks across
different platforms.
\subsection{Ensemble Toolkit}\label{ssec:entk}

An ensemble-based application is a \textbf{workflow}, i.e. tasks
with dependencies that determine the order of their execution. Subsets of
these tasks can be \textbf{workloads}, i.e., tasks whose dependencies have
been satisfied at a particular time and may be executed concurrently.
Ensemble-based application vary in the type of coupling between tasks, the
frequency and volume of information exchanged between these tasks, and the
executable of each task. This type of applications requires specific
coordination, orchestration and execution protocols, posing both
domain-specific and engineering challenges.

Ensemble Toolkit (EnTK), the topmost layer of RCT, simplifies the process of
creating and executing ensemble-based applications with complex coordination
and communication requirements. EnTK decouples the description of
ensemble-based applications from their execution by separating three orders
of concern: specification of task and resource requirements; resource
selection and acquisition; and task execution management. Domain scientists
retain full control of the implementation of their algorithms, programming
ensemble-based applications by describing what, when and where should be
executed. EnTK uses a runtime system, like RADICAL-Pilot, to acquire the
resources needed by applications to manage task execution.

\begin{figure}
  \centering
  \includegraphics[width=\columnwidth]{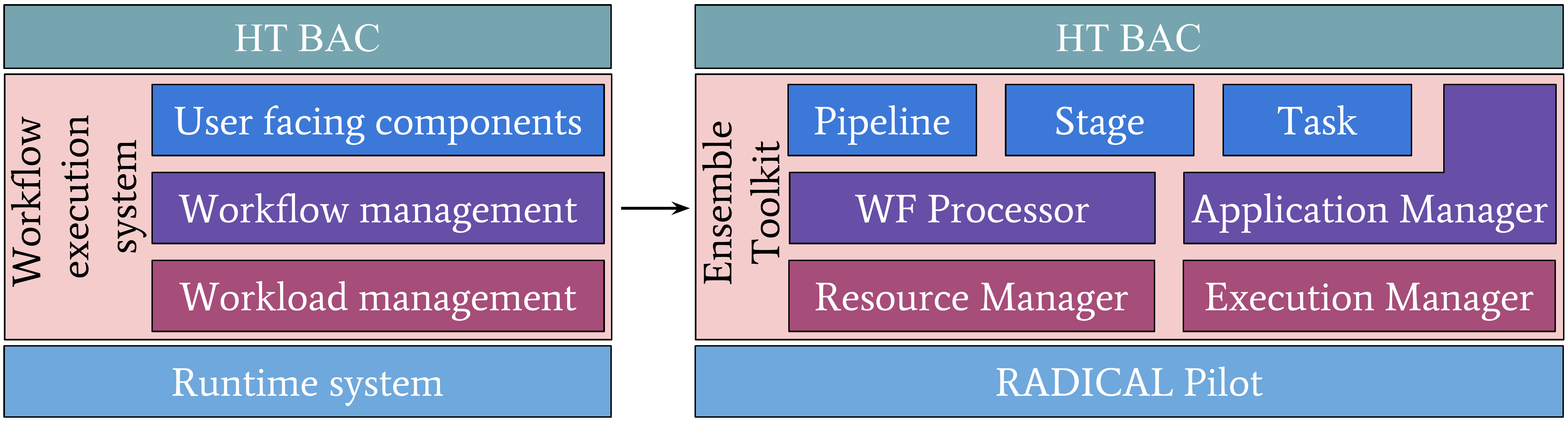}
  \caption{\textbf{Left:} Ensemble Toolkit overview showing how the abstract
  workflow execution system is mapped to specific components exposed
  to the users and components internal to the toolkit.}\label{fig:entk_arch}
\end{figure}

EnTK enables the creation of ensemble-based applications by exposing an 
application programming interface (API)
with four components: \textbf{Application Manager}, \textbf{Pipeline},
\textbf{Stage} and \textbf{Task}. Users describe ensembles in terms of
pipelines, stages and tasks, and pass this description to an instance
of Application Manager, specifying what resource to use for executing the
application (see Figure~\ref{fig:entk_arch}).

The Task component is used to encapsulate an executable and its software
environment and data dependencies. The Stage component contains a set of
tasks without mutual dependencies and that can therefore be executed
concurrently. The Pipeline component is used to describe a sequence of
stages, i.e., sets of tasks that need to be executed sequentially, not
concurrently.

The use of the Task, Stage, and Pipeline components, implemented as set and
list data structures, avoids the need to express explicitly relationship
among tasks. These relationships are insured by design, depending on the
formal properties of the lists and sets used to partition tasks into stages
and group stages into pipelines. Further, EnTK enables an explicit definition
of pre and post conditions on the execution of tasks, enabling a fine grained
adaptivity, both a local and global level. Conveniently, this does not
require the codification of a directed acyclic graph (DAG), a process that
imposes a rigid representation model on the domain
scientists~\cite{balasubramanian2017powerofmany}.


The \textbf{Application Manager} component of EnTK enables users to specify
target resources for the execution of the ensemble-based application. This
includes properties like walltime, number of nodes and credentials for
resource access. Users can also define execution setup parameters such as the
number of processes or messaging queues that should be used by EnTK\@. This
allows to size and tune the performance of EnTK, depending on the number of
tasks, stages and pipelines, but also on the resources available to the
toolkit.

The Application Manager along with the \textbf{WF Processor} is responsible
for the transformation of the application workflow into workloads, i.e., set
of tasks, that can be submitted to the indicated resources for execution.
Internally, the \textbf{Resource Manager} and \textbf{Execution Manager}
components enable the acquisition of resources and the management of
execution of these workloads (see \textbf{Figure~\ref{fig:entk_arch}}).

\section{High Throughput Binding Affinity Calculator}\label{sec:htbac}
Initially, we designed HTBAC to implement a single binding affinity protocol,
using the EnTK programming model to express the application logic. Here, we
exclusively focus on ESMACS to capture the workflow logic and isolate the
performance of a single protocol instance. HTBAC has been extended as a
Python library that enables the selection of multiple protocol instances of
ESMACS and TIES~\cite{dakka}.

A simulation pipeline is a defined sequence of simulation stages for a given
physical system. In the ESMACS protocol, these simulation pipelines are
replicated, where replicas differ only by their parameter configurations,
namely initial velocities, which are randomly generated and assigned by NAMD
at the start of execution. A simulation pipeline in the ESMACS protocol has 7
stages: the first, second and last stages perform staging of the input/output
data, the middle stages indicate simulation tasks. A task is appended to a
stage and stages are appended to a pipeline to maintain temporal order during
execution.



\begin{figure}
\centering
  \includegraphics[width=0.4\textwidth]{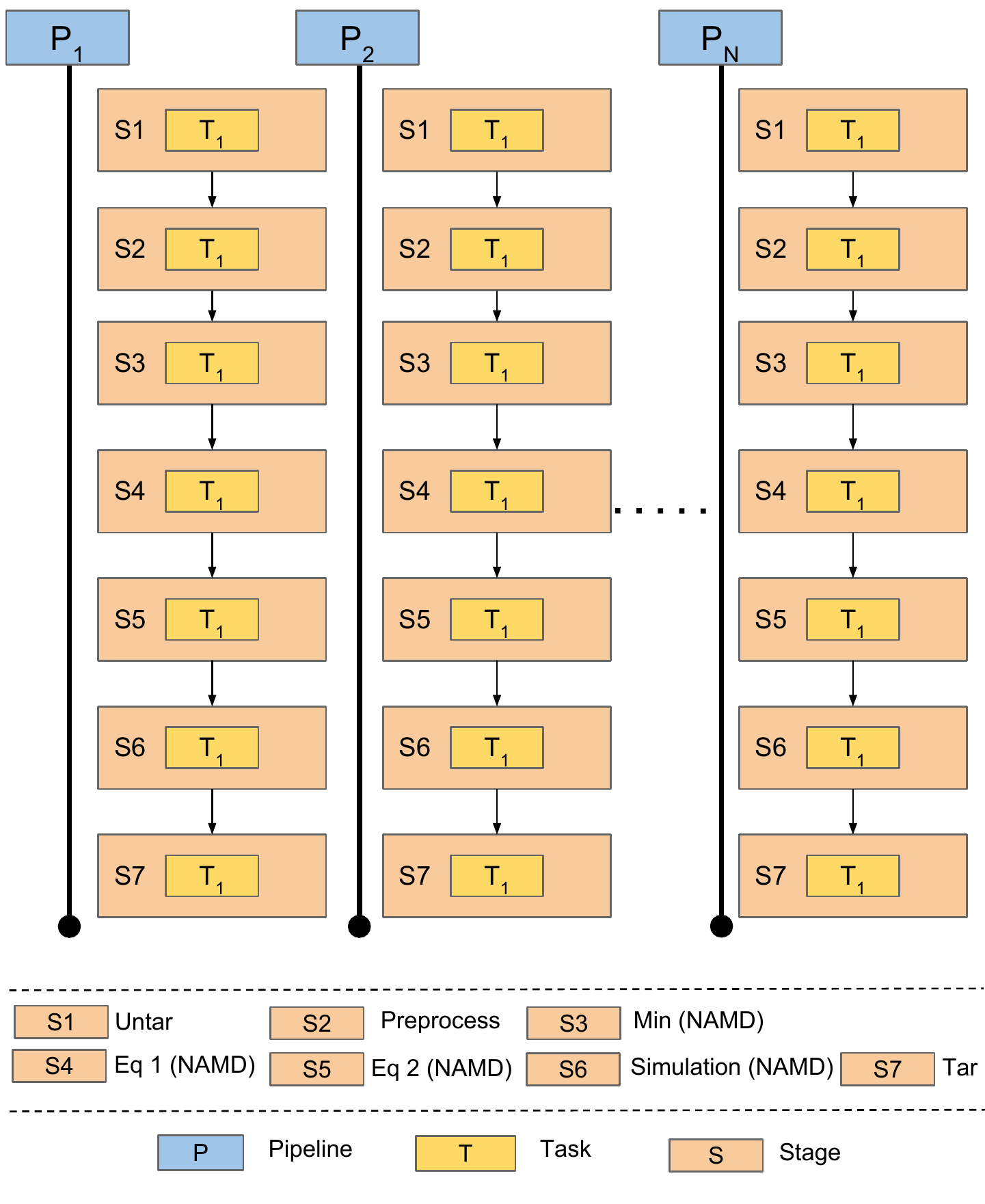}
  \caption{ESMACS protocol implemented as an ensemble application, encoded
  using the EnTK API\@. A protocol represents a physical system and is
  encoded as a set of independent pipelines. Each pipeline maps to a single
  replica. ESMACS consists of 25 replicas. Stages within a pipeline are
  executed sequentially. Each stage contain a single task performing unique
  functions, as required by the protocol. Stages S3--S6 contain molecular
  dynamics simulation tasks executed with NAMD}
  \label{figure:HTBAC}
\end{figure}

Each simulation pipeline replica maps to an independent EnTK pipeline. Each
pipeline consists of a sequence of stages, and each stage consists of a single
task that performs unique functions, including pre-processing and molecular
dynamics simulations. Fig~\ref{figure:HTBAC} shows how pipelines, stages and
tasks are organized for the ESMACS protocol. A task is composed of a set of
attributes that define parameters like the location of input files, the number
of simulations and the MD engine(s) used to launch those simulations.


\begin{figure}
\centering
  \includegraphics[width=0.5\textwidth]{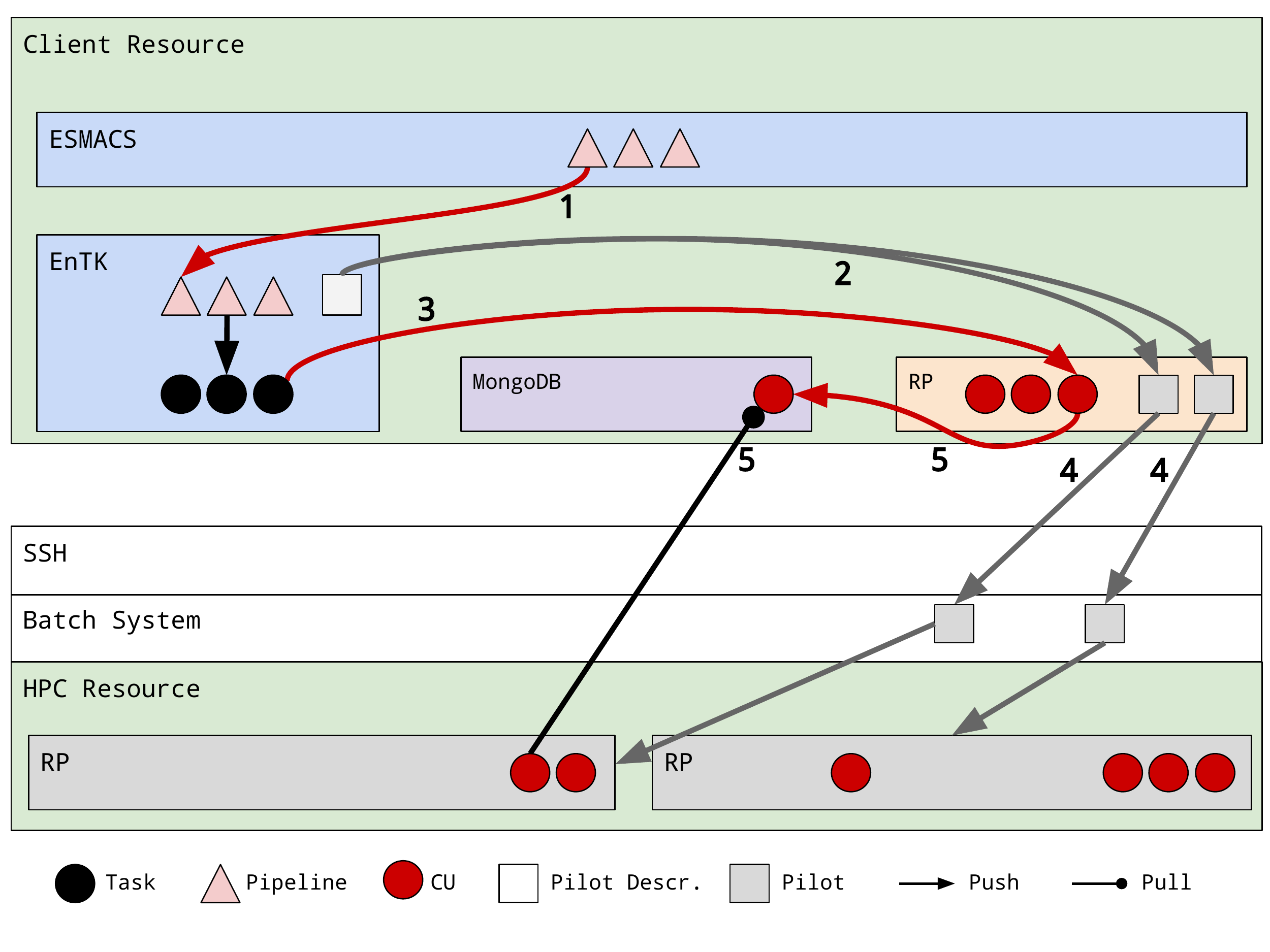}
  \caption{ESMACS-EnTK-RP Integration.
  Integration between ESMACS and EnTK\@. Numbers indicate
  the temporal sequence of execution. The database (DB) of RADICAL-Pilot (RP)
  can be deployed on any host reachable from the resources. RP pushes compute
  units (CU) to DB and pulls them for execution.}
  \label{figure:ht-bac_rp}
\end{figure}

Fig.~\ref{figure:ht-bac_rp} shows how the ESMACS protocol integrates with
EnTK\@. EnTK converts the set of pipelines into a set of tasks called compute
unit descriptions and submits them to RP\@. In addition, EnTK provides
methods for the user to specify a resource request including walltime, cores,
queue, and user credentials. EnTK converts this resource request into a pilot
that RP submits to a HPC machine. Once the pilot becomes active, it pulls
compute unit descriptions in bulk from a database, executing them on the
pilot resources.

\section{HTBAC Experiments}\label{sec:6}

Before embarking on a computational campaign that will consume 150M core
hours on the NCSA Blue Waters machine, we studied the scalability of HTBAC so
as to determine optimal workflow sizing and resource utilization for the
ESMACS protocol. The goal is twofold: (1) understanding the invariance of
HTBAC execution time over the number of workflow pipelines executed; and (2)
studying how the performance of EnTK and RP varies in relation to the size of
workflow.

\subsection{Experiment Design}\label{ssec:exp_design}

We designed two experiments to measure HTBAC, EnTK and RP weak scalability
when executing an increasing number of concurrent pipelines. According to the
use case described in Section~\ref{sec:htbac}, each pipeline consists of
seven stages, each stage with a single task. EnTK manages the queuing of the
tasks in accordance with the order and concurrency mandated by stages and
pipelines: For each pipeline, each stage is executed sequentially while
pipelines are executed concurrently.

Experiment 1 measures the baseline behavior of EnTK and RP with the workflow
of the ESMACS protocol and a null workload (\textmd{/bin/sleep 0}). The goal
is to isolate the overheads of EnTK and RP from the specifics of the
executables of the workflow and the overheads of the resources. The null
workload does not require data staging, I/O on both memory and disk, or
communication over network.

Experiment 2 replicates the design of Experiment 1 but it executes the
workflow of the ESMACS protocol with the actual simulation and data for the
EGFR kinase workload. The comparison between the two experiments enables
performance analysis of EnTK and RP to understand whether and how the size of
the executed workflow affects its overheads. Further, Experiment 2 shows also
whether HTBAC execution time is sensitive to the number of concurrent
pipelines executed.

Both experiments measure the weak scalability of HTBAC, EnTK and RP\@. This
means that the ratio between the number of pipelines and cores is kept
constant by design. While an investigation of strong scalability would
contribute to a better understanding of the behavior of HTBAC, EnTK and RP,
it is of limited interest for the current use case. The driving goal of HTBAC
is to increase throughput by a means of concurrency of pipelines, not in the
number of sequential executions per core. This is a driving motivation to
target large HPC machines instead of so-called high-throughput computing (HTC) 
infrastructures.

\subsection{Experiment Setup}\label{ssec:exp_setup}

We perform both Experiment 1 and 2 on NCSA's Blue Waters---a 13.3 petaFLOPS
Cray, with 32 Interlago cores/50 GB RAM per node, Cray Gemini, Lustre shared
file system. Currently, we exclusively use CPUs on Blue Waters as GPUs are
not required by our use case. RCT support the use of both type of
architectures and we previously benchmarked the use of GPUs.

We perform our experiments from a virtual machine hosted in Europe. This
helps to simulate the conditions in which the experimental campaign will be
performed by the research group at UCL\@. This is relevant because, as most
HPC resources, Blue Waters does not allow for executing applications on the
login node of the cluster. To this end, RCTs support \textmd{gsissh} for X509
authentication and authorization.

Table~\ref{tab:exp} shows the setup for Experiment 1 and 2. The ESMACS
protocol is executed with up to 25 concurrent but independent pipelines and
therefore their concurrent execution does not entail communication overhead.
Further, the EGFR kinase studies can benefit from greater concurrency because
potential HTBAC users may wish to extend their protocols beyond the current
scale of ESMACS\@. Consistently, our experiments push the boundaries of
current scale by executing 8, 16, 32, 64 and 128 concurrent pipelines.


\begin{table*}[t]
\centering
\caption{Experiment 1 executes the 7 stages of the ESMACS protocol with
a null workload; Experiment 2 uses the actual MD workload of the ESMACS
protocol. ESMACS protocol with EGFR kinase workload: (1) Untar configuration
files; (2) Preprep; (3) Minimize with decreasing restraints; (4)
Equilibration: NVT simulation at 50K, with restraints; (5) Equilibration: NPT
simulation at 300K, with decreasing restraints; (6) Equilibration: NPT at
300k, no constraints; (7) Tarball output files.}\label{tab:exp}
\begin{tabular}{llllllll}
\toprule
\textbf{Experiment ID}      &
\textbf{Protocol}           &
\textbf{Workload}           &
\textbf{\# Trials}          &
\textbf{\# Pipelines}       &
\textbf{\# Stages}          &
\textbf{\# Tasks}           &
\textbf{\# Cores per pilot} \\
\toprule
1                           &
ESMACS                      &
Null workload               &
2                           &
8, 16, 32, 64, 128          &
7                           &
7                           &
64, 128, 256, 512, 1024     \\
2                           &
ESMACS                      &
EGFR kinase Workload        &
2                           &
8, 16, 32, 64, 128          &
7                           &
7                           &
64, 128, 256, 512, 1024     \\
\bottomrule
\end{tabular}
\end{table*}

EnTK uses RP to acquire resources via a single pilot. The size of the pilot
is contingent upon characterization of performance, in this case, weak
scalability. Accordingly, we request the maximum number of cores required by
the workload as the number of cores in a pilot. We use between 64 and 1024
cores in Experiment 2 as the NAMD executable used in stages 3, 4, 5, and 6
requires 8 cores. Stages 1, 2 and 7 require instead 1 core. The null workload
of Experiment 1 requires only 1 core per stage but we request the same number
of cores as for Experiment 2 to be able to compare the overheads of both EnTK
and RP across experiments.

All experiments use EnTK version 0.4.7 and RP version 0.42. The MD engine
used is NAMD-MPI\@. The equilibration tasks of stage 4 and 6 are assigned
5000 timesteps while the task of stage 5 requires 55000 timesteps. We ran two
trials of both the null and MD workload at each pipeline configurations.

\subsection{Results}\label{ssec:exp_results}


\begin{figure}
  \centering
  \includegraphics[width=\columnwidth]{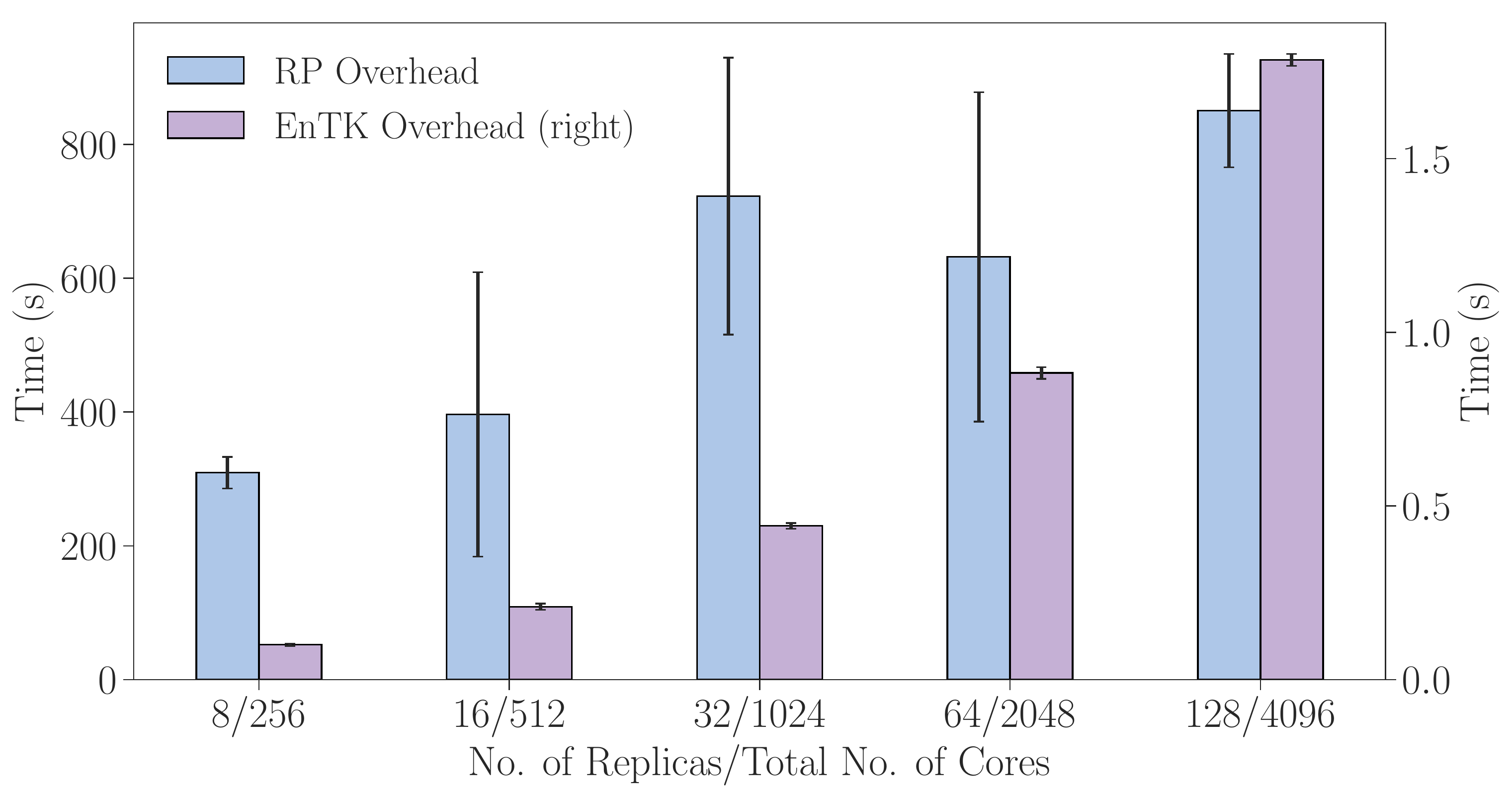}
  \caption{Overheads of Ensemble Toolkit (EnTK) and RADICAL-Pilot (RP) when
           executing HTBAC using a null workload. We plot the baseline
           EnTK/RP overheads without the application workload across two
           trials per pipeline configuration.}\label{fig:exp1}
\end{figure}

First we characterize the overhead of EnTK and RP in the null workload, where
we isolate the overhead introduced by the two systems
(Figure~\ref{fig:exp1}). We see a (slightly) superliner increase of EnTK
overhead, between 0.1 and 1.8 seconds. This overhead depends on the number of
tasks that need to be translated in-memory from a Python object to a CU 
description. As such, it is expected to grow proportionally to the
number of tasks, barring some competition of resources.

RP overhead is also, on average, superlinear but with a much greater
variance. This variance is due to mainly two factors: Network latency and
filesystem latency on the HPC resource. EnTK submits CU descriptions to the
MongoDB used by RP, and the RP pilot pulls these descriptions from the same
database. As described in Section~\ref{ssec:exp_setup}, this pull operation
occurs over a wide area network, introducing varying amounts of latency.
Further, RP writes and reads the CU descriptions multiple times to and
from the shared filesystem of the HPC machine. Together, these two factors
introduce delays in the scheduling of the CUs.


\begin{figure}
  \centering
  \includegraphics[width=\columnwidth]{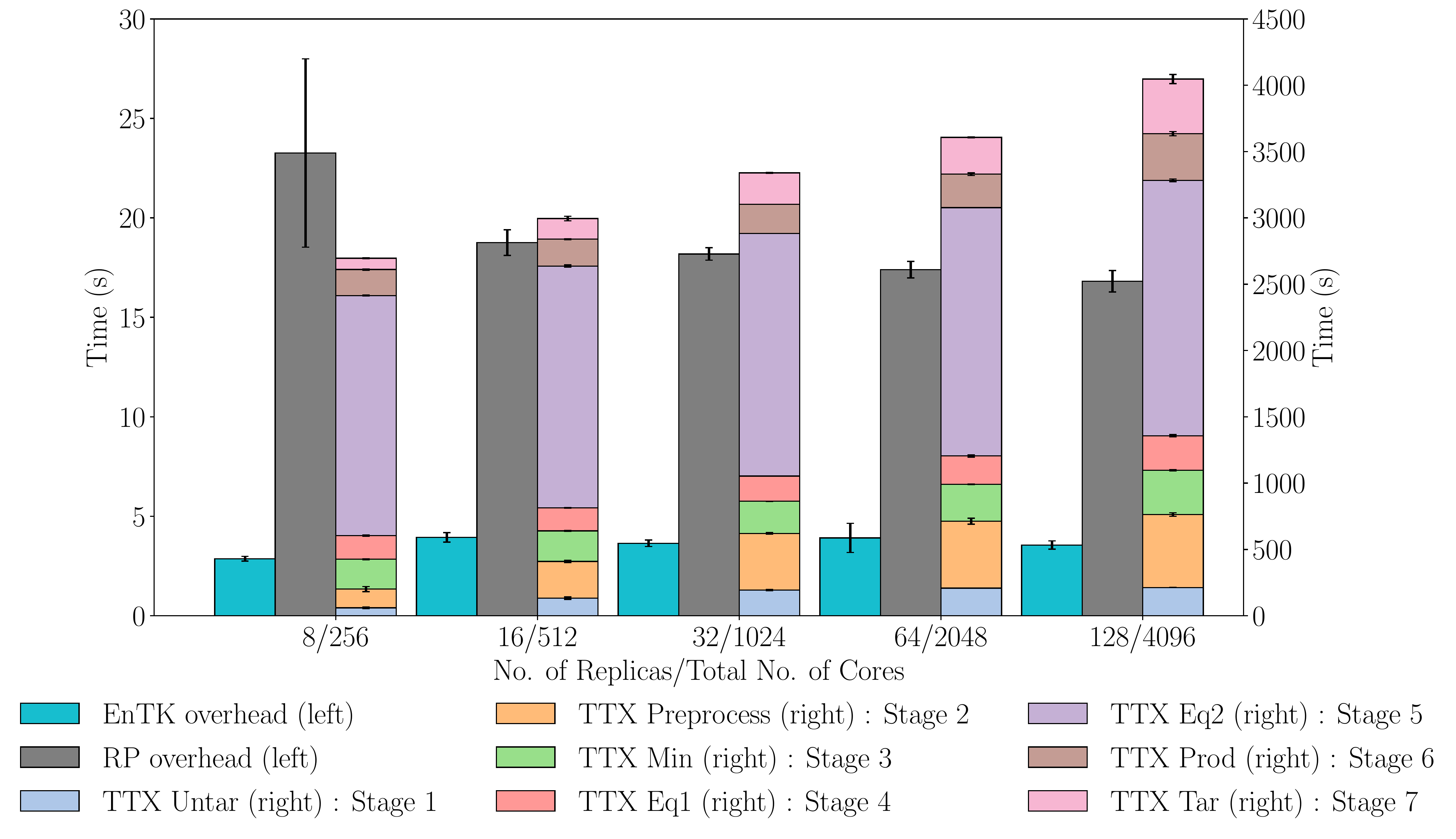}
  \caption{Similar EnTK/RP overhead behavior as with the null workload with higher
  values as the number of pipelines increases. We show a breakdown of TTX of
  each stage (Stage 1--7). Across pipeline configurations, TTX and RP
  overheads (accounting for the error bars) show weak scaling
  performance.}\label{fig:exp2}
\end{figure}

When the workload includes the EGFR kinase, we see (Figure~\ref{fig:exp2})
that the RP overhead becomes on average less than 10\% of the average total
execution time (TTX), defined as \(TTX = TTC - T_q\) where \(TTC\) is
time-to-completion and \(T_q\) is time spent queuing on the HPC machine. We
further break down TTX into the time-to-completion per stage, where stages
1,2, and 7 perform file movements, while stages 3,4,5, and 6 execute NAMD
tasks. At this level, we notice that the time-to-completion of the NAMD
stages are essentially invariant across pipelines of different size while
file movement stages exhibit linearly increasing behavior. In addition, when
accounting for variance, RP overheads also show linear weak scaling behavior.
As expected, EnTK overhead remains superlinear and comparable to the one
measured in Experiment 1. This is because in both experiments EnTK overhead
depends on the number of tasks translated to CU descriptions.

Experiments 1 and 2 show how the overheads of both EnTK and RP tend to be
invariant across type of workload executed. Their scaling behavior and, to
some approximation, their absolute values are comparable between
Figure~\ref{fig:exp1} and~\ref{fig:exp2}. This is relevant because it shows
that the systems used to coordinate and execute the ESMACS protocol add a
constant and comparatively not relevant overhead to the execution of NAMD\@.

\section{Discussion and Conclusion}\label{sec:conclusion}
It is necessary to move beyond the prevailing paradigm of running individual
MD simulations, which provide irreproducible results and cannot provide
meaningful error bars~\cite{Bhati2017}. Further, the ability to flexibly
scale and adapt ensemble-based protocols to the systems of interest is vital
to produce reliable and accurate results on timescales which make it viable
to influence real world decision making. To meet these goals, we are
designing and developing the high-throughput binding affinity calculator
(HTBAC).

HTBAC employs the RADICAL-Cybertools to build ensemble-based applications for
executing protocols like ESMACS at scale. We show how the implementation of
the ESMACS protocol scales almost perfectly to hundreds of concurrent
pipelines of binding affinity calculations on Blue Waters. 
This permits a time-to-solution that is essentially invariant of the size
of candidate ligands, as well as the type and number of protocols concurrently employed.



The use of software implementing well-defined abstractions like that of
``building blocks'', future proofs users of HTBAC to evolving hardware
platforms, while providing immediate benefits of scale and support for a
range of different application workflows. Thus, HTBAC represents an important
advance towards the use of molecular dynamics based free energy calculations
to the point where they can produce actionable results both in the clinic and
industrial drug discovery.

In the short term, the development of HTBAC will allow a significant increase
in the size of study. Much of the literature on MD-based free energy
calculations is limited to a few tens of systems, usually of similar drugs
bound to the same protein target. By facilitating investigations of much
larger datasets, HTBAC also provides a step towards tackling grand challenges
in drug design and precision medicine, where it is necessary to understand
the influences on binding strength for hundreds or thousands of drug-protein
variant combinations. Only in aiming to meet this ambitious goals we will be
able to reveal the limits of existing simulation technology and the
potentials used to approximate the chemistry of the real systems.


\section*{Acknowledgements}

\footnotesize

Access to Blue Waters was made possible by NSF OAC 1713749. The software
capabilities were supported by RADICAL-Cybertools (OAC 1440677) and NSF ICER
1639694. We thank Andre Merzky and other members of the RADICAL team for
support. PVC, SJZ and DWW would like to acknowledge the support of the EU
H2020 CompBioMed (675451), EUDAT2020 (654065) and ComPat (671564) projects. 
We acknowledge MRC Medical Bioinformatics project (MR/ L016311/1), 
and funding from the UCL Provost. 

{\it Author Roles and Contribution:} PVC and SJ conceived this work. MT and SJ
designed the experiments with support from JD and VB. JD performed the bulk of
experiments on Blue Waters, with support from DWW and VB. VB developed the
version of EnTK used for experiments. MT, VB, JD and SJ analyzed the data. SJZ
initially developed the ESMACS protocol using EnTK. DWW, MT and SJ wrote the
bulk of the paper.

HTBAC, Ensemble Toolkit and
RADICAL-Pilot can be found respectively at:
\url{https://github.com/radical-cybertools/htbac},
\url{https://github.com/radical-cybertools/radical.entk}, and
\url{https://github.com/radical-cybertools/radical.pilot}. Raw data and
scripts to reproduce experiments can be found at:
\url{https://github.com/radical-experiments/htbac-experiments.}






\bibliographystyle{plain}
\bibliography{rutgers,ucl}

\end{document}